\documentclass[a4paper,aps,prd,10pt,preprintnumbers,showpacs,showkeys,twocolumn,superscriptaddress,nofootinbib]{revtex4-1}
\usepackage{orcidlink,graphicx,cmap,amsmath}
\usepackage[utf8]{inputenc}
\usepackage[T1]{fontenc}
\usepackage{booktabs}
\usepackage{hyperref}
\usepackage{amssymb}

\def\re#1{Re(#1)}
\def\im#1{Im(#1)}
\def\Order#1{{\cal O}\left(#1\right)}

\begin{document}

\title{Quasinormal Modes and Grey-Body Factors of Scalar, Electromagnetic and Dirac Fields Around Einasto-Supported Regular Black Holes}
\author{S. V. Bolokhov \orcidlink{0000-0002-9533-530X}
}
\email{bolokhov-sv@rudn.ru}
\affiliation{RUDN University, 6 Miklukho-Maklaya St, Moscow, 117198, Russian Federation}

\begin{abstract}
We study quasinormal modes and grey-body factors of scalar, electromagnetic and Dirac test fields for a black hole surrounded by matter distributed according to the Einasto density profile. The quasinormal spectrum is calculated using the high-order WKB method with Padé approximants and checked by the time-domain integration. For small values of the Einasto index $\tilde n=1/2$ and $\tilde n=1$, the fundamental modes remain close to their Schwarzschild values, while for $\tilde n=5$ the oscillation frequency increases and the damping rate decreases as the halo parameter grows. Grey-body factors are much less sensitive to the halo environment: the main effect is a mild suppression at low frequencies caused by a moderate increase of the effective potential near the horizon. At higher frequencies the transmission probabilities remain close to the Schwarzschild case.
\end{abstract}


\maketitle

\section{Introduction}

In astrophysical environments black holes are typically embedded in extended matter distributions, most prominently galactic dark-matter halos. Observational evidence for dark matter is well established through galactic rotation curves, gravitational lensing, and large-scale structure formation. Although the microscopic nature of dark matter remains unknown, its macroscopic gravitational effects are accurately described by phenomenological density profiles such as the Navarro–Frenk–White, Hernquist, and Einasto models \cite{Navarro:1996gj,Bertone:2005xz,Hernquist:1990be,Einasto:1965}. These profiles provide effective descriptions of realistic halo structures without specifying the underlying particle physics.

The fundamental aspect of black-hole physics is related to the occurrence of spacetime singularities in classical general relativity which indicates that the theory ceases to be predictive in regimes of extremely strong curvature. According to the singularity theorems, under rather general assumptions gravitational collapse leads to geodesic incompleteness, implying that classical spacetime must be modified at sufficiently small scales. This expectation has motivated extensive efforts to construct black-hole solutions in which the central singularity is replaced by a regular core with finite curvature invariants. Understanding the properties of such regular geometries and determining whether they may represent realistic alternatives to classical black holes remain important open problems in gravitational physics.

Over the past decades a large variety of regular black-hole models has been proposed 
and studied  \cite{Bardeen:1968,Hayward:2005gi,Simpson:2018tsi,Ansoldi:2008jw,AyonBeato:1998ub,Dymnikova:1992ux,Bronnikov:2000vy,Bronnikov:2024izh,Bronnikov:2005gm,Kazakov:1993ha,Modesto:2008jz,Bonanno:2000ep,Lan:2023cvz,Konoplya:2023ppx,Bonanno:2023rzk,Bonanno:2025dry,Spina:2025wxb,Konoplya:2023ahd,Zhang:2024ney,Solodukhin:2025opw,Konoplya:2022hll,Bueno:2024dgm,Bueno:2025tli,Bueno:2024eig,Frolov:2026rcm,Konoplya:2020ibi,Bronnikov:2006fu,Bronnikov:2003gx,Casadio:2001jg,Konoplya:2024hfg}. 
Some constructions rely on phenomenological modifications of the metric function designed to regularize the central region while preserving asymptotic flatness. Other approaches interpret the regular core as supported by effective matter sources, often modeled through nonlinear electrodynamics or anisotropic fluids. In addition, a number of frameworks motivated by quantum gravity — such as asymptotically safe gravity, loop quantum gravity, or higher-curvature and nonlocal extensions of general relativity — naturally lead to regular geometries in which classical singularities are replaced by finite-curvature interiors. Although these scenarios differ in their physical motivation, they share the common feature that the strong-field region of the spacetime is modified while the geometry at large distances remains close to the Schwarzschild solution.

An important aspect of black-hole physics concerns the interaction between the spacetime geometry and propagating fields. In realistic astrophysical settings black holes are subject to external perturbations and interact with various forms of radiation and matter. The response of a black hole to perturbations is characterized by its quasinormal modes (QNMs), which describe damped oscillations determined solely by the properties of the background spacetime \cite{Kokkotas:1999bd,Konoplya:2011qq}. These modes dominate the ringdown stage of perturbed black holes and therefore play a key role in gravitational-wave observations. The quasinormal spectrum encodes information about the effective potential governing wave propagation in the vicinity of the photon sphere and near-horizon region, making it a sensitive probe of deviations from the Schwarzschild geometry.

Closely related to the quasinormal spectrum are the grey-body factors describing the scattering of waves by the gravitational potential barrier surrounding the event horizon. In contrast to QNMs, which correspond to purely outgoing solutions, grey-body factors characterize the transmission and reflection probabilities of waves incident on the black hole. These quantities determine the modification of the thermal Hawking radiation spectrum caused by spacetime curvature and thus control the energy emission rates of black holes \cite{Page:1976df,Page:1976ki,Kanti:2014vsa}. From a physical viewpoint, grey-body factors provide complementary information to quasinormal modes, as both are governed by the same effective potentials and reflect different aspects of wave propagation in curved spacetime.

Recently it has been shown that certain dark-matter distributions can be incorporated directly into exact black-hole solutions in such a way that the matter profile itself ensures regularity of the geometry. In particular, a family of regular black-hole metrics supported by an Einasto-type density distribution has been constructed in \cite{Konoplya:2025ect}. In this framework the same matter profile that successfully describes galactic halos also removes the central singularity, thereby establishing an interesting connection between astrophysical environments and regular black-hole geometries. Axial gravitational pertubrations of such Einasto-supported black holes have been recently investigated in \cite{Lutfuoglu:2026zel}, while spectra of Dehnen-supported profiles were studied in  \cite{Bolokhov:2025fto,Saka:2025xxl,Lutfuoglu:2025mqa}.

The purpose of the present work is to investigate the propagation of test fields in these regular black-hole backgrounds sourced by the Einasto distribution of matter. We analyze quasinormal modes and grey-body factors for massless scalar, electromagnetic, and Dirac fields propagating in the spacetime supported by the Einasto density profile. By studying the dependence of the quasinormal spectrum and transmission probabilities on the parameters characterizing the halo distribution, we determine how the presence of a realistic matter environment modifies the wave dynamics around regular black holes. Such an analysis allows one to assess the stability of these configurations, identify possible deviations from the Schwarzschild case, and clarify the interplay between strong-field geometry and astrophysical matter distributions.

The paper is organized as follows. In Sec.~\ref{sec:EinastoBackground} we review the regular black-hole solutions supported by the Einasto density profile and discuss their main geometric properties. In Sec.~\ref{sec:waveeq} we derive the perturbation equations for massless scalar, electromagnetic, and Dirac fields propagating in this background. Section~\ref{sec:methods} briefly outlines the methods used for calculating quasinormal modes and grey-body factors. In Secs.~\ref{sec:QNMs} and \ref{sec:gbf} we present the obtained spectra and transmission probabilities and analyze their dependence on the parameters of the Einasto distribution. Finally, Sec.~\ref{sec:conc} summarizes our results and discusses possible directions for further research.

\section{Regular black holes generated by Einasto matter distributions}
\label{sec:EinastoBackground}

We consider static and spherically symmetric spacetimes described by the line element
\begin{equation}
ds^2=-f(r)\,dt^2+\frac{dr^2}{f(r)}+r^2 d\sigma^2 .
\end{equation}
The geometry is conveniently parameterized through the mass function $m(r)$,
\begin{equation}
f(r)=1-\frac{2m(r)}{r}.
\end{equation}
The matter content supporting the configuration is modeled as an anisotropic effective fluid characterized by the energy density $\rho(r)$ and radial pressure $P_r(r)$. Following the construction proposed in~\cite{Konoplya:2025ect}, we impose the relation
\begin{equation}
P_r(r)=-\rho(r),
\end{equation}
which ensures the regular behavior of the metric at the horizon and leads to the simplified form of the metric function above. Under this assumption the Einstein equations reduce to a single equation determining the mass distribution,
\begin{equation}
m(r)=4\pi \int_0^r x^2 \rho(x)\,dx ,
\end{equation}
with the natural condition $m(0)=0$. If the density profile remains finite at the origin, the resulting geometry is regular and curvature scalars remain bounded everywhere.

\textbf{Einasto density distribution.} As a physically motivated example for the matter source we adopt the Einasto density profile,
\begin{equation}
\rho(r)=\rho_0\exp\!\left[-\left(\frac{r}{h}\right)^{1/\tilde{n}}\right],
\qquad \tilde{n}>0 ,
\end{equation}
which has become one of the most widely used empirical descriptions of dark-matter halos. Unlike earlier two-parameter models such as the Navarro--Frenk--White profile, the Einasto model introduces an additional shape parameter $\tilde{n}$ controlling the curvature of the density slope. Consequently the logarithmic derivative $d\ln\rho/d\ln r$ varies continuously with radius, allowing the profile to reproduce the gradual steepening of the density distribution observed in high-resolution cosmological simulations.

The Einasto model provides excellent fits to halos over a wide range of masses, from dwarf galaxies to galaxy clusters. In numerical simulations such as the Aquarius and Millennium projects, the density profiles of both host halos and subhalos are typically well approximated by Einasto functions with indices $\tilde{n}$ in the range $4\lesssim \tilde{n} \lesssim 8$. Observational studies based on stellar kinematics and rotation curves also favor Einasto-like distributions when reconstructing the dark-matter content of galaxies. For example, detailed modeling of spiral galaxies often yields $\tilde{n}\approx 5$--$7$, while clusters of galaxies tend to exhibit slightly larger values reflecting their different assembly histories.

The parameter $h$ defines the characteristic scale radius at which the logarithmic slope reaches a specific value, while $\rho_0$ sets the central density normalization. Integrating the density profile yields the total mass,
\begin{equation}
M=\lim_{r\rightarrow\infty} m(r),
\end{equation}
which is finite for any positive value of $\tilde{n}$. Consequently the spacetime approaches the Schwarzschild geometry at large distances. In the limit $h\rightarrow0$ with fixed $M$, the matter distribution collapses toward the origin and the metric continuously reduces to the Schwarzschild solution.

For arbitrary $\tilde{n}$ the integral determining $m(r)$ generally does not admit a simple elementary expression and must be evaluated numerically. Nevertheless, several particular values of $\tilde{n}$ allow analytic treatment and illustrate the essential properties of the resulting family of regular geometries.

\textbf{General properties.} For generic values of the shape parameter $\tilde{n}$ the spacetime exhibits a characteristic structure common to many regular black-hole models. Far from the central region the metric function approaches
\begin{equation}
f(r)\approx1-\frac{2M}{r},
\end{equation}
so that the gravitational field becomes indistinguishable from the Schwarzschild solution. Near the origin, however, the finite central density generates a de Sitter–like core whose curvature scale is determined by $\rho_0$.

The presence or absence of horizons depends on the compactness of the matter distribution. If the scale parameter $h$ is sufficiently small relative to the mass $M$, the spacetime contains two horizons corresponding to an outer event horizon and an inner Cauchy horizon. Increasing $h$ gradually decreases the compactness of the configuration, leading first to an extremal state where the two horizons coincide and eventually to a horizonless compact object.

The shape parameter $\tilde n$ determines how rapidly the density falls with radius and therefore controls the extent of the matter distribution. Smaller values of $\tilde{n}$ correspond to profiles that decay more sharply and produce geometries closer to Schwarzschild outside the central region. Larger values of $\tilde{n}$ lead to more extended halos, which produce stronger deviations from the vacuum metric in the vicinity of the horizon and typically reduce the parameter space admitting black-hole solutions.

Thus the Einasto-supported solutions interpolate smoothly between the Schwarzschild black hole and regular compact configurations with extended matter distributions. 

\textbf{Gaussian profile: $\tilde{n}=\tfrac12$.} A particularly simple case arises for $\tilde{n}=\tfrac12$, for which the density becomes Gaussian,
\begin{equation}
\rho(r)=\rho_0 \exp\!\left(-\frac{r^2}{h^2}\right).
\end{equation}
The corresponding mass function can be expressed in terms of the error function, yielding the metric function
\begin{equation}
f(r)=1-\frac{2M}{r}\,\mathrm{erf}\!\left(\frac{r}{h}\right)
+\frac{4M}{\sqrt{\pi}\,h}e^{-r^2/h^2},
\end{equation}
where the total mass is related to the central density through
\begin{equation}
M=\pi^{3/2}\rho_0 h^3 .
\end{equation}

Expanding the metric function near the center gives
\begin{equation}
f(r)=1-\frac{4M}{3\sqrt{\pi}h^3}\,r^2+\mathcal{O}(r^4),
\end{equation}
which corresponds to a de Sitter core with finite curvature. All curvature invariants remain bounded at $r=0$, confirming the absence of a central singularity. For sufficiently compact configurations (small $h/M$) the spacetime possesses two horizons. As the parameter $h$ increases, the horizons merge at a critical extremal configuration and disappear for larger values of $h$.

\textbf{Exponential profile: $\tilde{n}=1$.} Another analytically tractable example corresponds to $\tilde{n}=1$, leading to an exponential density distribution,
\begin{equation}
\rho(r)=\rho_0 e^{-r/h}.
\end{equation}
In this case the metric function becomes
\begin{equation}
f(r)=1-\frac{2M}{r}
+M\,\frac{2h^2+2hr+r^2}{h^2 r}e^{-r/h},
\end{equation}
where
\begin{equation}
M=8\pi \rho_0 h^3 .
\end{equation}

Near the center the expansion
\begin{equation}
f(r)=1-\frac{2M}{3h^3}r^2+\mathcal{O}(r^3)
\end{equation}
again reveals a regular de Sitter–type core. As in the Gaussian case, horizons exist only for sufficiently compact distributions. Because the exponential profile decreases more slowly with radius than the Gaussian one, the admissible range of $h/M$ leading to black-hole solutions is correspondingly narrower.

In the following sections we investigate the dynamical and observational properties of these geometries by studying wave propagation, quasinormal spectra and geodesic motion in the resulting spacetimes.

\begin{figure}
\resizebox{\linewidth}{!}{\includegraphics{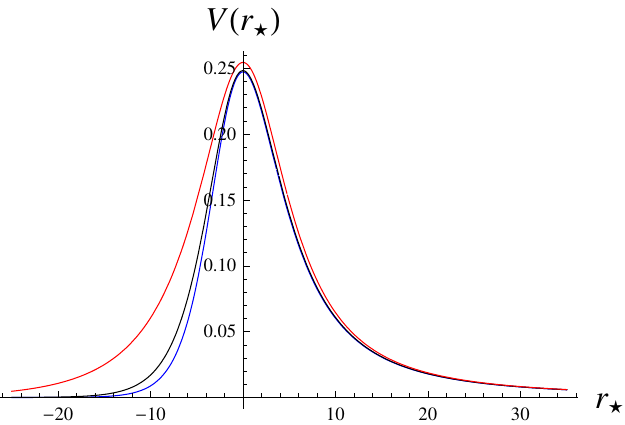}}
\caption{Typical effective potential as a function of the tortoise coordinate $r^{*}$ for small values of $\tilde n$. Here we have $\ell=0$ scalar perturbations; $M=1$; $h=0.1$ (blue), $h=0.3$ (black) and $h=0.38$ (red). The geometry is mostly corrected near the event horizon and quickly merge with the Schwarzschild case.}\label{fig:ScalarL0}
\end{figure}

\begin{table}
\begin{tabular}{c c c c}
\hline
$h$ & WKB16 ($\tilde{m}=8$) & WKB14 ($\tilde{m}=7$) & diff (\%) \\
\hline

\multicolumn{4}{c}{Scalar field, $\ell=0$} \\
\hline
$0.4$  & $0.110820-0.104502 i$ & $0.110679-0.104500 i$ & $0.0923$\\
$0.8$  & $0.112127-0.101107 i$ & $0.112443-0.101222 i$ & $0.223$\\
$0.85$ & $0.114945-0.102136 i$ & $0.112914-0.099867 i$ & $1.98$\\
$0.9$  & $0.113103-0.097845 i$ & $0.112755-0.097785 i$ & $0.236$\\
$0.95$ & $0.110906-0.095321 i$ & $0.109399-0.095113 i$ & $1.04$\\
$1.0$  & $0.112305-0.091485 i$ & $0.108554-0.079689 i$ & $8.54$\\
$1.05$ & $0.112009-0.085094 i$ & $0.111775-0.082471 i$ & $1.87$\\

\hline
\multicolumn{4}{c}{Scalar field, $\ell=1$} \\
\hline
$0.4$  & $0.27202$ & $0.292980-0.097680 i$ & $36.7$\\
$0.8$  & $0.292332-0.097046 i$ & $0.292321-0.097057 i$ & $0.0052$\\
$0.85$ & $0.291894-0.096459 i$ & $0.291914-0.096410 i$ & $0.0172$\\
$0.9$  & $0.291559-0.095480 i$ & $0.291361-0.095324 i$ & $0.0819$\\
$0.95$ & $0.290637-0.094078 i$ & $0.290621-0.093752 i$ & $0.107$\\
$1.0$  & $0.289585-0.092042 i$ & $0.289540-0.091769 i$ & $0.0912$\\
$1.05$ & $0.288295-0.089840 i$ & $0.288057-0.089118 i$ & $0.252$\\

\hline
\multicolumn{4}{c}{Electromagnetic field, $\ell=1$} \\
\hline
$0.4$ & $0.248263-0.092488 i$ & $0.248263-0.092488 i$ & $0$\\
$0.8$ & $0.247951-0.091482 i$ & $0.247986-0.091478 i$ & $0.0134$\\
$0.85$ & $0.248030-0.090736 i$ & $0.247822-0.090587 i$ & $0.0970$\\
$0.9$ & $0.247829-0.089563 i$ & $0.248168-0.089164 i$ & $0.198$\\
$0.95$ & $0.247515-0.087510 i$ & $0.247504-0.087451 i$ & $0.0229$\\
$1.0$ & $0.247165-0.084705 i$ & $0.247186-0.084771 i$ & $0.0266$\\
$1.05$ & $0.246405-0.081219 i$ & $0.246402-0.081177 i$ & $0.0160$\\

\hline
\multicolumn{4}{c}{Electromagnetic field, $\ell=2$} \\
\hline
$0.4$ & $0.457596-0.095004 i$ & $0.457596-0.095004 i$ & $0$\\
$0.8$ & $0.457156-0.094689 i$ & $0.457160-0.094685 i$ & $0.00129$\\
$0.85$ & $0.456862-0.094219 i$ & $0.456838-0.094229 i$ & $0.00539$\\
$0.9$ & $0.456478-0.093395 i$ & $0.456472-0.093363 i$ & $0.00686$\\
$0.95$ & $0.456024-0.092071 i$ & $0.456018-0.092071 i$ & $0.00132$\\
$1.0$ & $0.455515-0.090080 i$ & $0.455523-0.090078 i$ & $0.00162$\\
$1.05$ & $0.454906-0.087274 i$ & $0.454901-0.087274 i$ & $0.00107$\\

\hline
\multicolumn{4}{c}{Dirac field, $\ell=1/2$} \\
\hline
$0.4$ & $0.182856-0.096995 i$ & $0.182862-0.097149 i$ & $0.0741$\\
$0.8$ & $0.181849-0.095814 i$ & $0.181823-0.095826 i$ & $0.0140$\\
$0.85$ & $0.181614-0.094136 i$ & $0.169866-0.092481 i$ & $5.80$\\
$0.9$ & $0.181156-0.090928 i$ & $0.181540-0.090016 i$ & $0.488$\\
$0.95$ & $0.178839-0.089404 i$ & $0.179346-0.087949 i$ & $0.770$\\
$1.0$ & $0.173540-0.086944 i$ & $0.175657-0.084464 i$ & $1.68$\\
$1.05$ & $0.170594-0.082820 i$ & $0.171488-0.082139 i$ & $0.592$\\

\hline
\multicolumn{4}{c}{Dirac field, $\ell=3/2$} \\
\hline
$0.4$ & $0.380037-0.096405 i$ & $0.380037-0.096405 i$ & $0$\\
$0.8$ & $0.379341-0.096187 i$ & $0.379308-0.096201 i$ & $0.0094$\\
$0.85$ & $0.378821-0.095882 i$ & $0.378729-0.095751 i$ & $0.0411$\\
$0.9$ & $0.377986-0.094916 i$ & $0.377907-0.095040 i$ & $0.0378$\\
$0.95$ & $0.376886-0.093850 i$ & $0.376888-0.093858 i$ & $0.0021$\\
$1.0$ & $0.375568-0.092013 i$ & $0.375710-0.092482 i$ & $0.127$\\
$1.05$ & $0.373507-0.089630 i$ & $0.373365-0.089982 i$ & $0.0987$\\

\hline
\end{tabular}
\caption{Funamental QNMs of fields of various spin for $\tilde{n}=1/2$, $M=1$, calculated using WKB formulas of different orders with Padé approximants together with the difference between them.}
\end{table}

\begin{table}
\begin{tabular}{c c c c}
\hline
$h$ & WKB16 ($\tilde{m}=8$) & WKB14 ($\tilde{m}=7$) & diff (\%) \\
\hline

\multicolumn{4}{c}{$\ell=0$} \\
\hline
$0.1$ & $0.110447-0.104667 i$ & $0.110415-0.104700 i$ & $0.0304$\\
$0.15$ & $0.110539-0.104653 i$ & $0.110788-0.104613 i$ & $0.166$\\
$0.2$ & $0.110765-0.104230 i$ & $0.110765-0.104230 i$ & $0.0001$\\
$0.25$ & $0.110222-0.101514 i$ & $0.110217-0.101526 i$ & $0.0091$\\
$0.3$ & $0.111606-0.095913 i$ & $0.112729-0.095751 i$ & $0.771$\\
$0.35$ & $0.110363-0.089341 i$ & $0.111008-0.088995 i$ & $0.515$\\
$0.36$ & $0.109424-0.087883 i$ & $0.110316-0.087296 i$ & $0.761$\\
$0.37$ & $0.108130-0.086547 i$ & $0.109348-0.085641 i$ & $1.10$\\
$0.38$ & $0.106683-0.085479 i$ & $0.108233-0.084174 i$ & $1.48$\\

\hline
\multicolumn{4}{c}{$\ell=1$} \\
\hline
$0.1$ & $0.292936-0.097660 i$ & $0.292936-0.097660 i$ & $0.\times10^{-4}$\\
$0.15$ & $0.292923-0.097646 i$ & $0.292924-0.097646 i$ & $0.0001$\\
$0.2$ & $0.292837-0.097364 i$ & $0.292837-0.097364 i$ & $0.0002$\\
$0.25$ & $0.292843-0.096059 i$ & $0.292844-0.096055 i$ & $0.0015$\\
$0.3$ & $0.293524-0.092858 i$ & $0.293516-0.092870 i$ & $0.0047$\\
$0.35$ & $0.295431-0.086771 i$ & $0.295431-0.086771 i$ & $0$\\
$0.36$ & $0.295960-0.085055 i$ & $0.295962-0.085054 i$ & $0.0009$\\
$0.37$ & $0.296504-0.083132 i$ & $0.296511-0.083126 i$ & $0.0029$\\
$0.38$ & $0.297034-0.080986 i$ & $0.297048-0.080974 i$ & $0.0064$\\

\hline
\multicolumn{4}{c}{$\ell=2$} \\
\hline
$0.1$ & $0.483644-0.096759 i$ & $0.483644-0.096759 i$ & $0$\\
$0.15$ & $0.483634-0.096752 i$ & $0.483634-0.096752 i$ & $0$\\
$0.2$ & $0.483550-0.096550 i$ & $0.483550-0.096550 i$ & $0.00002$\\
$0.25$ & $0.483600-0.095440 i$ & $0.483600-0.095440 i$ & $0.00002$\\
$0.3$ & $0.484674-0.092475 i$ & $0.484674-0.092475 i$ & $0.00002$\\
$0.35$ & $0.488144-0.086540 i$ & $0.488144-0.086540 i$ & $0.00006$\\
$0.36$ & $0.489263-0.084825 i$ & $0.489263-0.084825 i$ & $0.00010$\\
$0.37$ & $0.490549-0.082867 i$ & $0.490550-0.082866 i$ & $0.00011$\\
$0.38$ & $0.492007-0.080618 i$ & $0.492007-0.080618 i$ & $0.00004$\\

\hline
\end{tabular}
\caption{Funamental QNMs of a scalar field for $\tilde{n}=1$, $M=1$, calculated using WKB formulas of different orders with Padé approximants together with the difference between them.}
\end{table}

\begin{table}
\begin{tabular}{c c c c}
\hline
$h$ & WKB16 ($\tilde{m}=8$) & WKB14 ($\tilde{m}=7$) & diff (\%) \\
\hline

\multicolumn{4}{c}{Electromagnetic field, $\ell=1$}\\
\hline
$0.1$ & $0.248263-0.092487 i$ & $0.248263-0.092488 i$ & $0.0003$\\
$0.15$ & $0.248258-0.092466 i$ & $0.248258-0.092466 i$ & $0.0002$\\
$0.2$ & $0.248294-0.092123 i$ & $0.248291-0.092124 i$ & $0.0012$\\
$0.25$ & $0.248825-0.090705 i$ & $0.248826-0.090705 i$ & $0.0004$\\
$0.3$ & $0.250707-0.087366 i$ & $0.250704-0.087371 i$ & $0.0021$\\
$0.35$ & $0.254782-0.080900 i$ & $0.254778-0.080914 i$ & $0.0055$\\
$0.36$ & $0.255887-0.079011 i$ & $0.255886-0.079025 i$ & $0.0053$\\
$0.37$ & $0.257054-0.076839 i$ & $0.257058-0.076850 i$ & $0.0046$\\
$0.38$ & $0.258240-0.074347 i$ & $0.258243-0.074347 i$ & $0.001$\\

\hline
\multicolumn{4}{c}{Electromagnetic field, $\ell=2$}\\
\hline
$0.1$ & $0.457595-0.095004 i$ & $0.457595-0.095004 i$ & $0$\\
$0.15$ & $0.457586-0.094995 i$ & $0.457586-0.094995 i$ & $0$\\
$0.2$ & $0.457531-0.094758 i$ & $0.457531-0.094758 i$ & $0$\\
$0.25$ & $0.457752-0.093554 i$ & $0.457752-0.093554 i$ & $0$\\
$0.3$ & $0.459290-0.090447 i$ & $0.459291-0.090447 i$ & $0.00004$\\
$0.35$ & $0.463698-0.084290 i$ & $0.463698-0.084290 i$ & $0.00010$\\
$0.36$ & $0.465087-0.082501 i$ & $0.465087-0.082500 i$ & $0.00009$\\
$0.37$ & $0.466679-0.080445 i$ & $0.466679-0.080445 i$ & $0.00008$\\
$0.38$ & $0.468481-0.078064 i$ & $0.468481-0.078064 i$ & $0.00002$\\

\hline
\multicolumn{4}{c}{Dirac field, $\ell=1/2$}\\
\hline
$0.1$ & $0.182838-0.097070 i$ & $0.182839-0.097069 i$ & $0.0005$\\
$0.15$ & $0.182770-0.097063 i$ & $0.182784-0.097008 i$ & $0.0271$\\
$0.2$ & $0.182591-0.096535 i$ & $0.182586-0.096533 i$ & $0.0025$\\
$0.25$ & $0.182222-0.094800 i$ & $0.182242-0.094760 i$ & $0.0217$\\
$0.3$ & $0.182231-0.091182 i$ & $0.181973-0.090274 i$ & $0.463$\\
$0.35$ & $0.182566-0.084573 i$ & $0.181991-0.084405 i$ & $0.298$\\
$0.36$ & $0.182680-0.082604 i$ & $0.182226-0.082493 i$ & $0.233$\\
$0.37$ & $0.182723-0.080388 i$ & $0.182349-0.080313 i$ & $0.191$\\
$0.38$ & $0.182603-0.077927 i$ & $0.182286-0.077881 i$ & $0.162$\\

\hline
\multicolumn{4}{c}{Dirac field, $\ell=3/2$}\\
\hline
$0.1$ & $0.380037-0.096405 i$ & $0.380037-0.096405 i$ & $0.00007$\\
$0.15$ & $0.380020-0.096399 i$ & $0.380020-0.096399 i$ & $0$\\
$0.2$ & $0.379859-0.096185 i$ & $0.379859-0.096184 i$ & $0.\times10^{-4}$\\
$0.25$ & $0.379649-0.094995 i$ & $0.379651-0.094999 i$ & $0.0012$\\
$0.3$ & $0.380167-0.091863 i$ & $0.380169-0.091853 i$ & $0.00249$\\
$0.35$ & $0.382529-0.085663 i$ & $0.382513-0.085667 i$ & $0.0043$\\
$0.36$ & $0.383312-0.083872 i$ & $0.383301-0.083876 i$ & $0.0030$\\
$0.37$ & $0.384201-0.081829 i$ & $0.384193-0.081833 i$ & $0.0024$\\
$0.38$ & $0.385182-0.079491 i$ & $0.385175-0.079496 i$ & $0.0023$\\

\hline
\end{tabular}
\caption{Funamental QNMs of electromagnetic and Dirac fields for $\tilde{n}=1$, $M=1$, calculated using WKB formulas of different orders with Padé approximants together with the difference between them.}
\end{table}

\begin{table*}
\begin{tabular*}{\linewidth}{@{\extracolsep{\fill}}c c c c}
\hline
$h$ & $\omega$ ($\ell=0$) & $\omega$ ($\ell=1$) & $\omega$ ($\ell=2$) \\
\hline
$0.1\times10^{-6}$   & $0.112162-0.104392 i$ & $0.293087-0.0975303 i$ & $0.483871-0.0966365 i$ \\
$0.2\times10^{-6}$   & $0.112831-0.103596 i$ & $0.294294-0.0969963 i$ & $0.485738-0.0961099 i$ \\
$0.3\times10^{-6}$   & $0.113882-0.102681 i$ & $0.296873-0.0963068 i$ & $0.489819-0.0954379 i$ \\
$0.4\times10^{-6}$   & $0.115362-0.101788 i$ & $0.300702-0.0956457 i$ & $0.495974-0.0948005 i$ \\
$0.5\times10^{-6}$   & $0.117199-0.100963 i$ & $0.305658-0.0950762 i$ & $0.504026-0.0942601 i$ \\
$0.6\times10^{-6}$   & $0.119341-0.100220 i$ & $0.311670-0.0946126 i$ & $0.513870-0.0938301 i$ \\
$0.7\times10^{-6}$   & $0.121763-0.0995598 i$ & $0.318721-0.0942498 i$ & $0.525483-0.0935049 i$ \\
$0.8\times10^{-6}$   & $0.124453-0.0989706 i$ & $0.326834-0.0939738 i$ & $0.538912-0.0932701 i$ \\
$0.9\times10^{-6}$   & $0.127416-0.0984360 i$ & $0.336080-0.0937658 i$ & $0.554278-0.0931064 i$ \\
$1.0\times10^{-6}$ & $0.130663-0.0979343 i$ & $0.346568-0.0936022 i$ & $0.571778-0.0929902 i$ \\
$1.1\times10^{-6}$ & $0.134211-0.0974380 i$ & $0.358463-0.0934524 i$ & $0.591697-0.0928907 i$ \\
$1.2\times10^{-6}$ & $0.138087-0.0969116 i$ & $0.371996-0.0932750 i$ & $0.614440-0.0927663 i$ \\
$1.3\times10^{-6}$ & $0.142318-0.0963134 i$ & $0.387491-0.0930089 i$ & $0.640580-0.0925556 i$ \\
$1.4\times10^{-6}$ & $0.146939-0.0956167 i$ & $0.405414-0.0925580 i$ & $0.670940-0.0921613 i$ \\
$1.5\times10^{-6}$ & $0.151902-0.0961122 i$ & $0.426453-0.0917564 i$ & $0.706754-0.0914149 i$ \\
\hline
\end{tabular*}
\caption{Quasinormal frequencies $\omega$ as functions of the parameter $\mu$ for scalar perturbations with multipole numbers $\ell=0,1,2$. BH model with $\tilde{n}=5$.}
\end{table*}

\begin{table*}
\begin{tabular*}{\linewidth}{@{\extracolsep{\fill}}c c c c}
\hline
$h$ & $\omega$ ($\ell=1$) & $\omega$ ($\ell=2$) & $\omega$ ($\ell=3$) \\
\hline
$0.1\times10^{-6}$  & $0.248462-0.092374 i$ & $0.457843-0.0948841 i$ & $0.657220-0.0954963 i$ \\
$0.2\times10^{-6}$  & $0.249864-0.0919406 i$ & $0.459805-0.0943840 i$ & $0.659838-0.0949854 i$ \\
$0.3\times10^{-6}$  & $0.252694-0.0914347 i$ & $0.464013-0.0937653 i$ & $0.665551-0.0943433 i$ \\
$0.4\times10^{-6}$  & $0.256772-0.0909991 i$ & $0.470296-0.0931975 i$ & $0.674163-0.0937458 i$ \\
$0.5\times10^{-6}$  & $0.261951-0.0906741 i$ & $0.478462-0.0927359 i$ & $0.685428-0.0932520 i$ \\
$0.6\times10^{-6}$  & $0.268151-0.0904691 i$ & $0.488404-0.0923904 i$ & $0.699203-0.0928733 i$ \\
$0.7\times10^{-6}$  & $0.275370-0.0903677 i$ & $0.500100-0.0921526 i$ & $0.715459-0.0926025 i$ \\
$0.8\times10^{-6}$  & $0.283629-0.0903540 i$ & $0.513603-0.0920075 i$ & $0.734270-0.0924248 i$ \\
$0.9\times10^{-6}$  & $0.293006-0.0904103 i$ & $0.529037-0.0919356 i$ & $0.755810-0.0923206 i$ \\
$1.0\times10^{-6}$ & $0.303624-0.0905133 i$ & $0.546605-0.0919130 i$ & $0.780361-0.0922661 i$ \\
$1.1\times10^{-6}$ & $0.315661-0.0906333 i$ & $0.566601-0.0919093 i$ & $0.808337-0.0922307 i$ \\
$1.2\times10^{-6}$ & $0.329368-0.0907295 i$ & $0.589444-0.0918828 i$ & $0.840320-0.0921726 i$ \\
$1.3\times10^{-6}$ & $0.345098-0.0907410 i$ & $0.615721-0.0917721 i$ & $0.877135-0.0920304 i$ \\
$1.4\times10^{-6}$ & $0.363360-0.0905699 i$ & $0.646281-0.0914795 i$ & $0.919973-0.0917063 i$ \\
$1.5\times10^{-6}$ & $0.384914-0.0900443 i$ & $0.682401-0.0908341 i$ & $0.970624-0.0910300 i$ \\
\hline
\end{tabular*}
\caption{Quasinormal frequencies $\omega$ for electromagnetic perturbations with multipole numbers $\ell=1,2,3$ as functions of the parameter $h$ for the model with $\tilde{n}=5$.}
\end{table*}

\begin{figure*}
\resizebox{\linewidth}{!}{\includegraphics{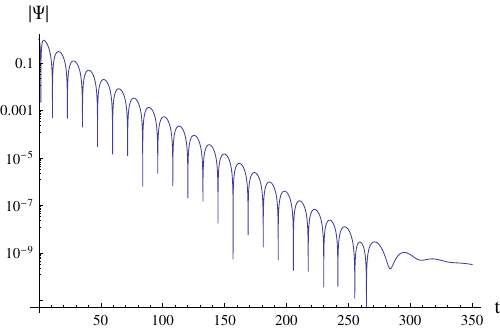}\includegraphics{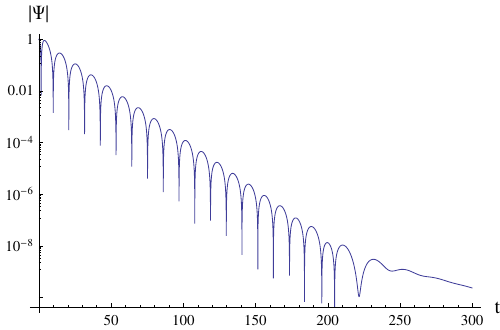}}
\caption{Semi-logarithmic time-domain profiles for $\ell=1$ electromagnetic perturbations of the black hole model with $\tilde{n}=1$, $h=0.38$ and or $\ell=1$ scalar perturbations of the black hole model with $\tilde{n}=1/2$ $h=1.05$. The Prony method allows to extract the dominant frequency $\omega = 0.288190 - 0.0895577 i$, which is very close to the WKB data $\omega = 0.288295 - 0.089840 i$.}\label{fig:TDMaxwellL1}
\end{figure*}

\section{Perturbations of test fields}
\label{sec:waveeq}

In order to investigate the dynamical response of the spacetime we consider test perturbations of scalar, electromagnetic, and Dirac fields propagating in the background geometry
\begin{equation}
ds^2=-f(r)\,dt^2+\frac{dr^2}{f(r)}+r^2(d\theta^2+\sin^2\theta\,d\phi^2).
\end{equation}
For all three types of perturbations the field equations can be reduced, after separation of variables (see, for instance \cite{Brill:1957fx,Moncrief:1974gw,Konoplya:2018arm}), to a Schrödinger-like wave equation of the form
\begin{equation}
\frac{d^2\Psi}{dr_*^2}+\left(\omega^2-V(r)\right)\Psi=0,
\label{masterwave}
\end{equation}
where the tortoise coordinate $r_*$ is defined through
\begin{equation}
\frac{dr_*}{dr}=\frac{1}{f(r)} .
\end{equation}
The effective potential $V(r)$ depends on the spin of the perturbing field and determines the spectrum of quasinormal modes.

\textbf{Scalar field.} We first consider a test scalar field $\Phi$ obeying the Klein–Gordon equation
\begin{equation}
\square \Phi=0 ,
\end{equation}.
Using the standard separation of variables
\begin{equation}
\Phi(t,r,\theta,\phi)=e^{-i\omega t}Y_{\ell m}(\theta,\phi)\frac{\Psi(r)}{r},
\end{equation}
with $Y_{\ell m}$ the spherical harmonics, one obtains a radial equation for $\Psi(r)$. Introducing the tortoise coordinate reduces the equation to the form (\ref{masterwave}) with the effective potential
\begin{equation}
V_s(r)=f(r)\left(\frac{\ell(\ell+1)}{r^2}+\frac{f'(r)}{r}\right).
\label{scalarpotential}
\end{equation}

The first term corresponds to the angular momentum barrier, the second term arises from spacetime curvature, and the last term reflects the presence of the scalar-field mass.

\textbf{Electromagnetic field.} Electromagnetic perturbations are governed by Maxwell's equations
\begin{equation}
\nabla_\mu F^{\mu\nu}=0,
\end{equation}
where $F_{\mu\nu}$ is the electromagnetic field tensor. Expanding the vector potential in vector spherical harmonics and separating variables leads to a single master equation for the radial function.

After transforming to the tortoise coordinate the equation again takes the form (\ref{masterwave}) with the effective potential
\begin{equation}
V_{em}(r)=f(r)\frac{\ell(\ell+1)}{r^2}.
\label{empotential}
\end{equation}

Unlike the scalar case, the electromagnetic potential contains only the centrifugal barrier multiplied by the metric function.

\textbf{Dirac field.} Finally we consider a massless Dirac field satisfying
\begin{equation}
\gamma^a e_a^{\ \mu}(\partial_\mu+\Gamma_\mu)\Psi_D=0,
\end{equation}
where $e_a^{\ \mu}$ denotes the tetrad and $\Gamma_\mu$ the spin connection.

Following the standard Chandrasekhar separation procedure, the Dirac equation reduces to a pair of coupled first-order radial equations. These can be combined into two decoupled second-order equations of the form (\ref{masterwave}) with supersymmetric partner potentials
\begin{equation}
V_{D\pm}(r)=W^2(r)\pm\frac{dW}{dr_*},
\end{equation}
where
\begin{equation}
W(r)=\frac{\kappa\sqrt{f(r)}}{r}, \qquad 
\kappa=\ell+\frac{1}{2} .
\end{equation}

Explicitly, the potentials can be written as
\begin{equation}
V_{D\pm}(r)=
f(r)\frac{\kappa^2}{r^2}
\pm
\frac{\kappa\sqrt{f(r)}}{r}
\left(
\frac{f'(r)}{2}-\frac{f(r)}{r}
\right).
\label{diracpotential}
\end{equation}

The two potentials $V_{D+}$ and $V_{D-}$ correspond to different chiralities of the Dirac field but are known to produce identical quasinormal spectra.

For asymptotically flat black holes the effective potentials for all fields vanish at spatial infinity and near the event horizon, as illustrated in fig. \ref{fig:ScalarL0}. Thus the potentials typically form a single barrier outside the event horizon. This structure makes the WKB approximation particularly suitable for computing the quasinormal spectrum.

\section{Methods for calculating quasinormal modes}
\label{sec:methods}

\textbf{Quasinormal-mode boundary conditions.} Quasinormal modes (QNMs) correspond to solutions of the wave equation
(\ref{masterwave}) subject to physically motivated boundary conditions \cite{Kokkotas:1999bd, Konoplya:2011qq, Berti:2009kk, Bolokhov:2025uxz}. For asymptotically
flat black-hole spacetimes the effective potential vanishes both near
the event horizon and at spatial infinity. Therefore the asymptotic
solutions take the form of free waves.

The QNM boundary conditions require purely ingoing waves at the event
horizon and purely outgoing waves at spatial infinity,
\begin{equation}
\Psi \propto
\begin{cases}
e^{-i\omega r_*}, & r_* \to -\infty, \\
e^{+i\omega r_*}, & r_* \to +\infty .
\end{cases}
\end{equation}

These conditions define a discrete set of complex frequencies
\begin{equation}
\omega = \omega_R + i\omega_I ,
\end{equation}
where $\omega_R$ determines the oscillation frequency and
$\omega_I<0$ corresponds to the damping rate of the perturbation.

\textbf{WKB approximation with Padé resummation.} To compute the quasinormal spectrum we employ the semi-analytic
Wentzel–Kramers–Brillouin (WKB) approximation developed for
black-hole perturbations \cite{Iyer:1986np,Konoplya:2003ii,Matyjasek:2017psv,Matyjasek:2019eeu}. This method is based on matching
the WKB solutions of the wave equation across the maximum
of the effective potential barrier.

In this approach the quasinormal frequencies satisfy the condition
\begin{equation}
\frac{i(\omega^2-V_0)}{\sqrt{-2V_0''}}
-
\sum_{j=2}^{N}\Lambda_j
=
n+\frac12 ,
\end{equation}
where $V_0$ is the value of the effective potential at its
maximum, $V_0''$ is the second derivative of the potential
with respect to the tortoise coordinate evaluated at the
same point, $n$ denotes the overtone number, and
$\Lambda_j$ represent higher-order WKB correction terms.

The WKB expansion can be carried to high orders, significantly
improving the accuracy of the method. In the present work we
use high-order WKB formulas supplemented by Padé
resummation, which accelerates the convergence of the
asymptotic series. The Padé transformation effectively
reconstructs the series in the form of a rational function,
allowing one to obtain stable results even when the original
WKB expansion converges slowly.

The WKB method is particularly efficient for modes with
moderate or large multipole number $\ell$ and for the
fundamental or low overtones ($n<\ell$), where the effective
potential has a single well-defined maximum. 

In the present work we employ the 14th- and 16th-order WKB formulas \cite{Matyjasek:2017psv,Matyjasek:2019eeu} for analytically tractable cases, and the 8th-order approximation for numerically constructed backgrounds. The structure of Padé
approximants $P^{\tilde m}_{\tilde n}$ is chosen such that $\tilde{m} = \tilde{n}$, where $\tilde{m}+\tilde{n}$ is equal to the WKB order \cite{Konoplya:2019hlu} (here $\tilde m$ and $\tilde n$ are the parameters of Padé approximation).  

The WKB approach has been thoroughly elaborated and widely employed in studies of black-hole perturbations and related wave-propagation problems. Numerous applications and further developments of the method can be found throughout the literature (see, for instance, \cite{Konoplya:2009hv,Malik:2024cgb,Bolokhov:2024ixe,Skvortsova:2023zmj,Han:2026fpn,Konoplya:2006ar,Bolokhov:2023dxq,Lutfuoglu:2025hjy,Konoplya:2025hgp,Malik:2025erb,Kanti:2006ua,Lutfuoglu:2026xlo,Arbelaez:2026eaz,Konoplya:2005sy,Malik:2024tuf,Stuchlik:2025ezz,Bolokhov:2023bwm,Skvortsova:2024atk,Konoplya:2023moy,Bolokhov:2024bke,Kodama:2009bf,Lutfuoglu:2025eik,Lutfuoglu:2025pzi} for representative examples). For this reason, we refrain from repeating the detailed derivation of the method here.

\textbf{Time-domain integration.} As an independent verification of the obtained frequencies
we also analyze the evolution of perturbations in the
time domain. For this purpose we integrate the wave
equation using the characteristic discretization scheme
introduced by Gundlach, Price, and Pullin.

Introducing null coordinates
\begin{equation}
u=t-r_*, \qquad v=t+r_* ,
\end{equation}
the wave equation takes the form
\begin{equation}
4\frac{\partial^2\Psi}{\partial u\,\partial v}
+V(r)\Psi=0 .
\end{equation}

The equation is solved numerically on a discretized
$(u,v)$ grid using the finite-difference relation
\begin{equation}
\Psi_N=\Psi_W+\Psi_E-\Psi_S
-\frac{\Delta^2}{8}V_S(\Psi_W+\Psi_E)
+\mathcal{O}(\Delta^4),
\end{equation}
where the subscripts denote the points of the numerical
grid.

The resulting time-domain profile typically consists
of three stages: an initial transient determined by the
chosen initial data, an intermediate phase dominated by
exponentially damped oscillations corresponding to the
quasinormal modes, and finally the late-time power-law
tail.

The dominant quasinormal frequency can be extracted
from the intermediate stage of the signal using fitting
techniques such as the Prony method. The time-domain integration approach has been widely applied in studies of black-hole perturbations, providing a reliable way to extract the dominant quasinormal frequency and to diagnose possible dynamical instabilities. Numerous works have demonstrated the effectiveness of this method in analyzing the temporal evolution of perturbations \cite{Konoplya:2005et,Skvortsova:2023zca,Dubinsky:2025wns,Konoplya:2013sba,Lutfuoglu:2025blw,Abdalla:2005hu,Dubinsky:2024gwo,Ishihara:2008re,Skvortsova:2025cah,Dubinsky:2024mwd,Lutfuoglu:2025mqa,Konoplya:2014lha,Cuyubamba:2016cug,Lutfuoglu:2025bsf,Dubinsky:2025bvf,Stuchlik:2025mjj}.

\section{Quasinormal modes}\label{sec:QNMs}

The numerical results presented in Tables~I--V allow one to assess the influence of the Einasto halo on the quasinormal spectrum of test fields. In particular, the tables contain both the computed quasinormal frequencies and a comparison of different orders of the WKB approximation, which provides an estimate of the numerical accuracy of the method.

For relatively small values of the Einasto index, $\tilde n=1/2$ and $\tilde n=1$, the fundamental quasinormal mode remains close to its Schwarzschild value. This can be seen from Tables~I--III, where the real part of the frequency undergoes only a small shift as the halo parameter $h$ varies, while the imaginary part, determining the damping rate, changes even less. For example, for the scalar field with $\ell=0$ the frequencies remain close to $\omega\approx 0.11-0.10 i$ throughout the considered range of $h$, with only moderate variations in both the oscillation frequency and the damping rate. A similar behavior is observed for higher multipoles. For $\ell=1$ the real part of the scalar-field frequency stays near $\omega_R\approx0.29$, while the imaginary part remains close to $\omega_I\approx-0.097$. For $\ell=2$ the frequencies are located near $\omega_R\approx0.48$ and $\omega_I\approx-0.097$, again showing only mild changes as the halo parameter varies. Thus the oscillation period and damping time remain close to the Schwarzschild values when $\tilde n$ is small.

The tables also demonstrate that the WKB approximation is well convergent in this regime. For instance, in Table~I the scalar-field frequencies are computed using the 16th-order WKB approximation with the Padé parameter $\tilde m=8$ and compared with the 14th-order WKB approximation with $\tilde m=7$. For $\ell=0$ the difference between these two orders typically remains at the level of $10^{-3}$--$10^{-1}\%$ for most values of $h$, increasing only near the boundary of the parameter range where the potential barrier becomes less suitable for the WKB treatment. A similar level of agreement is observed for $\ell=1$ and $\ell=2$, where the difference between the two WKB orders is typically well below one percent. 

The same pattern is visible for electromagnetic perturbations in Table~III. For example, for $\ell=1$ the difference between the 16th- and 14th-order WKB results remains extremely small, often at the level of $10^{-4}$--$10^{-3}\%$, even when the halo parameter varies significantly. For higher multipoles the agreement is similarly good. These comparisons indicate that the numerical uncertainty associated with the WKB approximation is very small and typically much smaller than the deviations caused by the presence of the halo.

A more pronounced modification of the spectrum appears when the Einasto index becomes larger. This is illustrated in Tables~IV and V for $\tilde n=5$. In this case the quasinormal frequencies change much more significantly as the parameter $h$ increases. For example, for the scalar field the real part of the frequency grows substantially with increasing $h$, while the magnitude of the imaginary part gradually decreases, indicating slower damping of perturbations. Similar trends are observed for higher multipoles: for $\ell=1$ and $\ell=2$ the oscillation frequency increases steadily across the table, while the damping rate changes more moderately. Thus the effect of the halo on the quasinormal spectrum becomes considerably stronger in this regime.

The difference between the cases $\tilde n=1/2,1$ and $\tilde n=5$ can be understood from the structure of the metric function generated by the Einasto profile. When $\tilde n$ is small, the spacetime differs from the Schwarzschild geometry mainly in a relatively narrow region near the event horizon. As a result, the effective potential governing perturbations remains close to the Schwarzschild potential in the region where the quasinormal modes are formed. In contrast, for $\tilde n=5$ the modification of the metric extends over a much larger radial interval, producing a more substantial change in the effective potential barrier and therefore in the resulting quasinormal frequencies.

Finally, the above conclusions are further supported by the time-domain profiles of perturbations (see Figs.~\ref{fig:TDMaxwellL1}), which confirm that the oscillation frequencies and damping rates extracted from the signal agree with the values obtained using the WKB method with great accuracy. 

Overall, the numerical data show that Einasto halos with small values of the parameter $\tilde n$ produce only weak modifications of the quasinormal spectrum, while larger values of $\tilde n$ lead to noticeably stronger deviations from the Schwarzschild case.
\begin{figure*}
\resizebox{\linewidth}{!}{\includegraphics{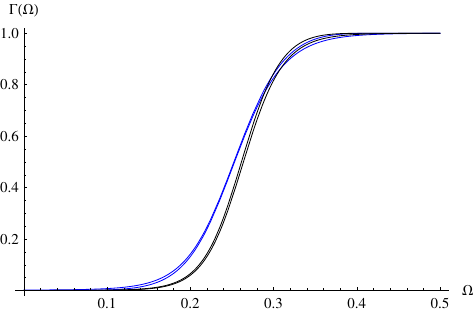}\includegraphics{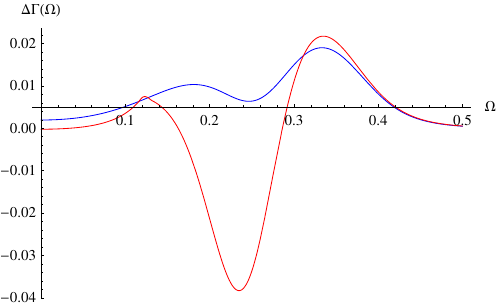}}
\caption{Left: Grey-body factors found by the 6th order WKB formula and by the correspondence with quasinormal modes. Right: Difference between the grey-body factors obtained by the two methods. Here we consider electromagnetic perturbations of the $\tilde{n}=1$ black-hole model at $\ell=1$, $h=0.01$ (red), and $h=0.38$ (black).}\label{fig:emL1}
\end{figure*}

\begin{figure*}
\resizebox{\linewidth}{!}{\includegraphics{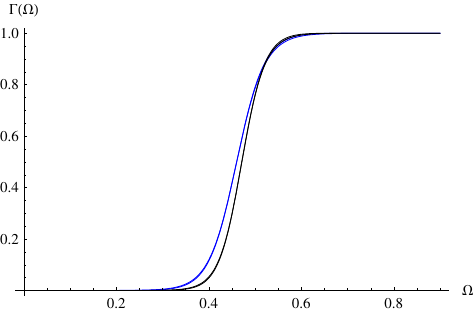}\includegraphics{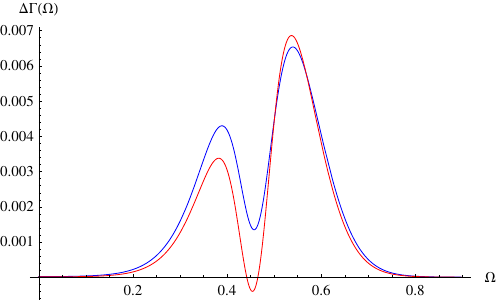}}
\caption{Left: Grey-body factors found by the 6th order WKB formula and by the correspondence with quasinormal modes. Right: Difference between the grey-body factors obtained by the two methods. Here we consider electromagnetic perturbations of the $\tilde{n}=1$ black-hole model at $\ell=2$, $h=0.01$ (red), and $h=0.38$ (black).}\label{fig:emL2}
\end{figure*}

\begin{figure*}
\resizebox{\linewidth}{!}{\includegraphics{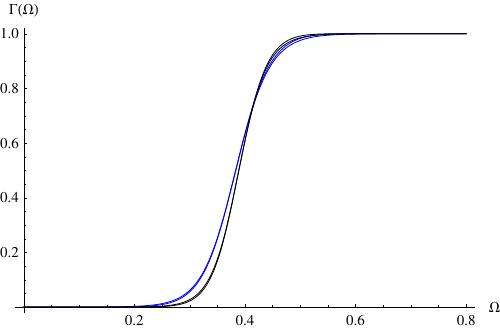}\includegraphics{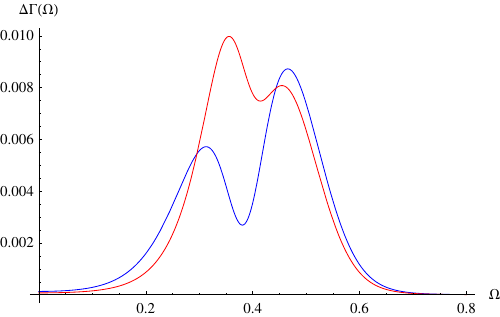}}
\caption{Left: Grey-body factors found by the 6th order WKB formula and by the correspondence with quasinormal modes. Right: Difference between the grey-body factors obtained by the two methods. Here we consider Dirac perturbations of the $\tilde{n}=1$ black-hole model at $\ell=3/2$, $h=0.01$ (red), and $h=0.38$ (black).}\label{fig:DiracL32}
\end{figure*}

\section{Grey-body factors}
\label{sec:gbf}

In addition to quasinormal modes, an important characteristic of wave
propagation in black-hole spacetimes is provided by grey-body factors,
which describe the probability that radiation emitted near the horizon
propagates to spatial infinity. These factors quantify the deviation
from the ideal blackbody spectrum of Hawking radiation caused by the
frequency-dependent scattering of waves by the effective potential
barrier surrounding the black hole.

For test fields propagating in a static spherically symmetric
background, the perturbation equations reduce to the Schrödinger-like
form \ref{masterwave}.

The scattering problem is defined by imposing a purely ingoing wave at the event horizon,
\begin{equation}
\Psi = T e^{-i\Omega r_*},
\qquad
r_* \to -\infty,
\end{equation}
while at spatial infinity the solution consists of a superposition of incoming and reflected waves,
\begin{equation}
\Psi = e^{-i\Omega r_*}+R\,e^{i\Omega r_*},
\qquad
r_* \to +\infty ,
\end{equation}
where $T$ is the transmission coefficient and $R$ is the reflection coefficient.

Conservation of the Wronskian implies
\begin{equation}
|R|^2+|T|^2=1 .
\end{equation}

The grey-body factor is defined as the transmission probability
\begin{equation}
\Gamma_\ell(\Omega)\equiv|T|^2 ,
\end{equation}
which determines the fraction of radiation of frequency $\Omega$ and multipole number $\ell$ that escapes to infinity.

The transmission coefficient can be estimated using the WKB method,
which is particularly effective when the effective potential has the
form of a single barrier. In this approach the transmission probability
is given by
\begin{equation}
|T|^2=
\left(
1+\exp\left[2\pi K\right]
\right)^{-1},
\end{equation}
where
\begin{equation}
K=\frac{\Omega^2-V_0}{\sqrt{-2V_0''}}
+
\sum_{i=2}^{N}\Lambda_i .
\end{equation}

Here $V_0$ is the maximum of the effective potential, $V_0''$ denotes its second derivative with respect to the tortoise coordinate at the maximum, and $\Lambda_i$ represent higher-order WKB correction terms. 

In the high--multipole (eikonal) regime the scattering problem for black-hole perturbations admits a particularly simple analytic description. In this limit the transmission probability through the effective potential barrier is directly
related to the fundamental quasinormal frequency of the system. The grey–body factor for multipole number $\ell$ can therefore be approximated as
\cite{Konoplya:2024vuj,Konoplya:2024lir}
\begin{equation}\label{transmission-eikonal}
\Gamma_{\ell}(\Omega)=
\left(1+e^{2\pi\dfrac{\Omega^2-\re{\omega_0}^2}{4\re{\omega_0}\im{\omega_0}}}\right)^{-1}
 + \Order{\frac{1}{\ell}}.
\end{equation}

The correction terms in the expansion (\ref{transmission-eikonal}) represent
systematic deviations from the strict eikonal limit. These higher–order
contributions contain additional information about the quasinormal spectrum,
including the influence of the first overtone, and provide successive
refinements of the analytic approximation. In practice, the accuracy of the
relation between the transmission probability and the quasinormal frequencies
has been explicitly verified beyond the leading geometric–optics order,
with calculations carried out up to the second order in the expansion beyond
the eikonal approximation.

The validity of this correspondence between grey–body factors and quasinormal
modes has been tested in numerous recent studies for a variety of black-hole
backgrounds and field perturbations
\cite{Dubinsky:2025nxv,Malik:2025erb,Bolokhov:2024otn,Lutfuoglu:2025ldc,
Han:2025cal,Malik:2024cgb,Lutfuoglu:2025hjy,Skvortsova:2024msa,
Bolokhov:2025lnt,Dubinsky:2024vbn,Lutfuoglu:2025ohb,Malik:2025dxn,
Lutfuoglu:2025blw}. These analyses demonstrate that the analytic relation
(\ref{transmission-eikonal}) reproduces the transmission coefficients with
high precision for sufficiently large multipole numbers. Neveretheless, there are cases when the WKB cannot effectively applied and consequently the correspondence does not work, as for example for a number of theories with higher curvature corrections where the eikonal limit of the effective potential does not have a standard centrifugal form \cite{Konoplya:2017ymp,Konoplya:2017zwo} or the matching of the Taylor series with the WKB expansions at the boundaries is inaccurate \cite{Bolokhov:2023dxq}.

The grey-body factors presented in Figs. \ref{fig:emL1} - \ref{fig:DiracL32} allow one to estimate the influence of the Einasto halo on the scattering properties of test fields. In contrast to the quasinormal spectrum discussed in the previous section, the grey-body factors appear to be considerably less sensitive to the presence of the halo environment.

From Fig. \ref{fig:emL1} - \ref{fig:DiracL32} one can see that the effective potentials governing the propagation of perturbations are only slightly modified when the halo parameter $h$ increases. In particular, the height of the potential barrier grows moderately in the region close to the event horizon. This change remains relatively small over the whole parameter range considered, and the overall shape of the potential barrier stays very similar to that of the Schwarzschild case. Since the grey-body factors are determined mainly by the transmission through this barrier, only a modest modification of the transmission probability is expected.

This behavior is indeed observed in the grey-body factors shown in Figs. \ref{fig:emL1} - \ref{fig:DiracL32}. For all types of perturbations considered, the curves corresponding to different values of the halo parameter remain close to each other across most of the frequency range. The deviations become noticeable mainly at low frequencies, where the transmission probability is the most sensitive to the precise height of the potential barrier. As the halo parameter $h$ increases and the potential barrier slightly grows near the horizon (Fig.~\ref{fig:ScalarL0}), the transmission probability at small frequencies becomes somewhat reduced. Consequently, the grey-body factors are mildly suppressed in the low-frequency regime.

At higher frequencies the influence of the halo becomes negligible. In this regime the transmission probability approaches unity and the grey-body factors corresponding to different values of the halo parameter practically coincide. This behavior is expected because high-frequency waves penetrate the potential barrier almost freely and therefore become insensitive to small changes in its shape.

Overall, the results demonstrate that the Einasto halo produces only a weak modification of the grey-body factors, even in the cases where noticeable changes in the quasinormal spectrum are present. In particular, while the quasinormal frequencies—especially higher overtones—can be sensitive to relatively small deformations of the effective potential, the grey-body factors depend mainly on the global properties of the barrier and therefore remain largely unaffected by the halo environment. The main visible effect of the halo is a slight suppression of the grey-body factors at low frequencies due to the moderate increase of the effective potential near the event horizon.

\vspace{5mm}
\section{Conclusion}\label{sec:conc}

A large body of literature is devoted to the study of black holes surrounded by galactic dark-matter halos, typically modeled within geometries that retain a central singularity. For an overview of such configurations see, for example, \cite{Bolokhov:2025zva} and the references cited therein. Perturbative properties of these spacetimes have been investigated in numerous works, particularly through the analysis of quasinormal modes, which characterize the response of the geometry to external disturbances \cite{Konoplya:2021ube,Chakraborty:2024gcr,Pezzella:2024tkf,Zhang:2021bdr,Liu:2024bfj,Feng:2025iao,Daghigh:2022pcr,Dubinsky:2025fwv,Konoplya:2022hbl,Zhao:2023tyo,Liu:2024xcd}. Closely related scattering characteristics, such as grey-body factors and the associated Hawking radiation spectra, have also been analyzed in a number of studies \cite{Mollicone:2024lxy,Tovar:2025apz,Lutfuoglu:2025kqp,Pathrikar:2025sin,Hamil:2025pte}.  Most of the models studied in this context assume that the spacetime possesses a central singularity.

We have studied the quasinormal modes and grey-body factors of scalar, electromagnetic and Dirac test fields for a regular black hole surrounded by matter distributed according to the Einasto density profile. The perturbation equations were reduced to wave-like forms with effective potentials and the quasinormal frequencies were calculated using the high-order WKB method with Padé approximants. Grey-body factors were obtained from the corresponding scattering problem.

The results show that the influence of the halo on the quasinormal spectrum depends strongly on the Einasto parameter $\tilde n$. For relatively small values of this parameter, $\tilde n=1/2$ and $\tilde n=1$, the fundamental quasinormal modes remain close to the Schwarzschild values. Both the oscillation frequency and the damping rate change only slightly as the halo parameter $h$ varies. In contrast, for $\tilde n=5$ the deviations become noticeably larger: the real part of the frequency increases with increasing $h$, while the magnitude of the imaginary part decreases, indicating slower damping of perturbations. This behavior reflects the fact that for small $\tilde n$ the spacetime differs from the Schwarzschild geometry mainly in a narrow region close to the horizon, whereas for larger $\tilde n$ the modification of the metric extends to a wider radial region and produces a more significant deformation of the effective potential barrier.

The comparison of different orders of the WKB approximation demonstrates good convergence of the method. The differences between the 16-th and 14th-order WKB results are typically very small and remain well below the overall change in the frequencies caused by the halo parameter, indicating that the observed effects are not related to numerical uncertainty.

The grey-body factors appear to be much less sensitive to the halo environment. The effective potentials change only slightly when the halo parameter increases, mainly through a moderate growth of the barrier close to the event horizon. As a consequence, the grey-body factors exhibit only small deviations from the Schwarzschild case. The main visible effect is a mild suppression of the transmission probability at low frequencies, while at higher frequencies the curves corresponding to different halo parameters practically coincide.

Overall, the Einasto halo produces only weak modifications of the scattering properties of test fields, while the quasinormal spectrum—especially for larger values of the Einasto index—shows a more noticeable response to the surrounding matter distribution.

\vspace{5mm}
\acknowledgments
I would like to thank R. A. Konoplya for useful discussions.

\bibliography{EinastoRegular}
\end{document}